\input epsf

\def\INSERTCAP#1#2{\vbox{%
{\narrower\noindent%
\multiply\baselineskip by 3%
\divide\baselineskip by 4%
{\rm Table #1. }{\sl #2 \medskip}}
}}
\input harvmac
\def\vcb{$|V_{cb}|$}
\def\CF{ {\cal F}}
\def\qmax{{q^2_{\rm max}}}
\def\OMIT#1{}
\def\ie{{\it i.e.}}

\def\spur{\raise.15ex\hbox{/}\kern-.57em }

\def\CO{{\cal O}}
\def\ccdot{\hbox{\kern-.1em$\cdot$\kern-.1em}}
\def\frac#1#2{{#1\over#2}}

\def\larr#1{\raise1.5ex\hbox{$\leftarrow$}\mkern-16.5mu #1}

%
%

%
\def\eps{\epsilon}
\def\bra#1{\left\langle #1\right|}
\def\ket#1{\left| #1\right\rangle}
%
%

\def\lsl{\spur {\kern0.1em l}}

%

%

\def\np#1#2#3{\NP{\bf B#1} (#2) #3}
\def\pl#1#2#3{\PL B {\bf #1} (#2) #3}
\def\prl#1#2#3{\PRL{\bf #1} (#2) #3}
\def\pr#1#2#3{\PR{\bf #1} (#2) #3}
\def\prep#1#2#3{\PRep{\bf #1} (#2) #3}

\def\sjnp#1#2#3{\SJNP{\bf #1} (#2) #3}

\def\NP{{Nucl.\ Phys.\ }}
\def\PL{{Phys.\ Lett.\ }}
\def\PR{{Phys.\ Rev.\ }}
\def\PRep{{Phys.\ Rep.\ }}
\def\PRL{{Phys.\ Rev.\ Lett.\ }}
\def\SJNP{{Sov.\ J. Nucl.\ Phys.\ }}
%
%
%
%
%
\def\INSERTFIG#1#2#3{\vbox{\vbox{\hfil\epsfbox{#1}\hfill}%
{\narrower\noindent%
\multiply\baselineskip by 3%
\divide\baselineskip by 4%
{\ninerm Figure #2. }{\ninesl #3 \medskip}}
}}%
%

%
\relax

%
\def\w{{\omega}}

\Title{\vbox{\hbox{UCSD/PTH 95-11}}}{\vbox{%
\centerline{Model-Independent Determinations of}
\centerline{$\bar B \to D l\bar \nu,D^* l\bar \nu$ Form Factors} }}
\centerline{C. Glenn Boyd\footnote{$^{\ast}$}{gboyd@ucsd.edu},
Benjam\'\i n Grinstein\footnote{$^{\dagger}$}{bgrinstein@ucsd.edu} and
Richard F. Lebed\footnote{$^{\ddagger}$}{rlebed@ucsd.edu}}
\bigskip\centerline{Department of Physics}
\centerline{University Of California, San Diego}
\centerline{La Jolla, California 92093-0319}
\vskip .3in
We present nonperturbative, model-independent parametrizations of the
individual QCD form factors relevant to $\bar B \to D^* l \bar \nu$
and $\bar B \to D l \bar \nu$ decays. These results follow from
dispersion relations and analyticity, without recourse to heavy quark
symmetry.  To describe a form factor with two percent accuracy, three
parameters are necessary, one of which is its normalization at zero
recoil, $\CF(1)$. We combine the individual form factors using heavy
quark symmetry to extract values for the product $|V_{cb}| \CF(1)$
from $\bar B \to D^* l \bar \nu$ data with negligible extrapolation
uncertainty.

%
\Date{August 1995} 
\nref\iw{N. Isgur and M.B. Wise, \pl{232}{1989}{113} and
\pl{237}{1990}{527}.}
\nref\eh{E. Eichten and B. Hill, \pl{234}{1990}{511}.}
\nref\vs{M. B. Voloshin and M. A. Shifman, Yad.\ Fiz.\ {\bf 47},
(1988) 801 [\sjnp{47}{1988}{511}].}
\nref\aleph{D. Buskulic {\it et al.\/} (ALEPH Collaboration), CERN
Report No.\ CERN-PPE/95-094 (unpublished).}
\nref\cleo{B. Barish {\it et al.\/} (CLEO Collaboration),
\pr{D51}{1995}{1014}.}
\nref\argus{H. Albrecht {\it et al.\/} (ARGUS Collaboration), Z.
Phys.\ C {\bf 57} (1993) 533.}
\nref\bglapr{C.G. Boyd, B. Grinstein, and R.F. Lebed, U.C. San Diego
Report No.\ UCSD/PTH 95-03 [hep-ph/9504235], to appear in Phys.\
Lett.\ B.}
\nref\thresh{A.F. Falk and M. Neubert, \pr{D47}{1993}{2695 and 2982};
M. Neubert, \pl{338}{1994}{84}; T. Mannel, \pr{D50}{1994}{428};
M. Shifman, N. Uraltsev, and A. Vainshtein, \pr{D51}{1995}{2217}.}
%
%
\nref\hist{N.N. Meiman, Sov.\ Phys.\ JETP {\bf 17} (1963) 830;
S. Okubo, \pr{D3}{1971}{2807}; S. Okubo and I. Fushih,
\pr{D4}{1971}{2020}; V. Singh and A.K. Raina, Fortschritte der Physik
{\bf 27} (1979) 561; C. Bourrely, B. Machet, and E. de Rafael,
\np{189}{1981}{157}; E. de Rafael and J. Taron, \pr{D50}{1994}{373}.}
\nref\dec{C.G. Boyd, B. Grinstein, and R.F. Lebed,
\prl{74}{1995}{4603}.}
\nref\klt{V.V. Kisilev, A.K. Likhoded, and A.V. Tkabladze,
\pr{D51}{1995}{3613}.}
\nref\eq{E.J. Eichten and C. Quigg, \pr{D49}{1994}{5845}.}
\nref\ck{Y.-Q. Chen and Y.-P. Kuang, \pr{D46}{1992}{1165}.}
\nref\blasref{P. Duren, {\it Theory of $H^p$ Spaces}, Academic Press,
New York, 1970.}
\nref\luke{M.E. Luke, \pl{252}{1990}{447}.}
\nref\hqben{B. Grinstein, in {\it Proceedings of the Workshop on B
Physics at Hadron Accelerators}, Snowmass, Colorado, 1993, ed.\
C.S. Mishra and P. McBride (Fermilab, 1994).}
\nref\neubrev{M. Neubert, \prep{245}{1994}{259}.}
\nref\pdg{L. Montanet {\it et al.\/} (Particle Data Group),
\pr{D50}{1994}{1173}.}
\nref\bjbd{J. Bjorken, in {\it Proceedings of the 4th Recontres de
Physique de la Vall\`{e}e d'Aoste}, La Thuille, Italy, 1990, ed.\
M. Greco (Editions Fronti\`{e}rs, Gif-Sur-Yvette, France, 1990);
N. Isgur and M.B. Wise, \pr{D43}{1991}{819}.}
\newsec{Introduction}

A determination of the Cabibbo-Kobayashi-Maskawa element \vcb\ from
the\break
 decays $\bar B \to D^* l \bar \nu$ and $\bar B \to D l \bar \nu$
requires knowledge of the transition amplitudes\break
$\vev{D^*(p',\epsilon)|(V_\mu-A_\mu)|\bar B(p)}$ and
$\vev{D(p')|V_\mu|\bar B(p)}$, respectively.  In the limit of
infinitely heavy $b$ and $c$ quark masses these amplitudes are
predicted\refs{\iw{--}\vs}\ at one kinematic point, namely, when the
recoiling $D^*$ or $D$ is at rest in the rest frame of the decaying
$\bar B$.  In terms of $q\equiv p-p'$, this zero recoil point occurs
at $q^2=\qmax\equiv(M-m)^2$, where $M=m_B$ is the decaying meson mass
and $m=m_D$ or $m_{D^*}$ is the mass of the final state meson.
Unfortunately, the differential decay width ${\rm d}\Gamma/{\rm d}q^2$
vanishes at $q^2_{\rm max}$, so an extraction of \vcb\ requires the
extrapolation of the matrix element from $q^2$ values less than
$\qmax$.

This method has been used by several
experiments\refs{\aleph{--}\argus}.  ARGUS\argus\ tested the
importance of the extrapolation on the determination of \vcb\ by using
various parametrizations. The observed variations of \vcb\ were larger
than the rest of the errors combined. In principle the error inherent
in the extrapolation can be made arbitrarily small by collecting an
arbitrarily large amount of data, arbitrarily close to $\qmax$; such
an endeavor is impractical, if not unattainable. Therefore, a
precision measurement of \vcb\ from $\bar B \to D l \bar \nu$ or $D^*
l \bar \nu$ requires a model-independent understanding of the
extrapolation.

In a previous letter\bglapr\ we presented such a model-independent
extrapolation. To this effect we used analyticity, crossing symmetry,
and QCD dispersion relations to find a two-parameter fit to the
$B$-meson $b$-number elastic form factor $F(q^2)$.  Heavy quark
symmetries were then invoked to relate $F(q^2)$ to the amplitude for
$\bar B \to D^* l \bar \nu$; given the validity of heavy quark
symmetries, we showed that over the relevant range of $q^2$ the
accuracy of the two-parameter fit was better than 1\%.

With a two-parameter extrapolation at hand, experimentalists can
accurately determine \vcb\ by making a simultaneous fit of the data to
\vcb\ and the two parameters in our extrapolation. At the moment this
program suffers from two main theoretical uncertainties:
%
\item{1~)}Incalculable nonperturbative corrections to the amplitudes
for $\bar B \to D l \bar \nu$ and $D^* l \bar \nu$ at $\qmax$ appear
at orders $1/m_c$ and $1/m_c^2$, respectively\thresh.  The size of
these is controversial.
\item{2~)}The extrapolation of Ref.~\bglapr\ relies on heavy quark
spin and flavor symmetries, with {\it a~priori} corrections of order
$1/m_c$.

Of these issues only the latter is addressed in this paper.  Instead
of assuming particular numerical values for the normalization of form
factors at zero recoil and making a fit to two parameters plus
$|V_{cb}|$, we evade the first issue by presenting our results as
three-parameter fits in units of $\CF(1)$, the amplitude at $\qmax$.
We then improve on the method of Ref.~\bglapr\ by dropping the
unnecessary use of heavy quark symmetries. To this end we derive, in
Sec.\ 2, bounds on the form factors describing the amplitudes for
$\bar B \to D l \bar \nu$ and $D^* l \bar \nu$. As in Ref.~\bglapr,
our arguments are based on QCD dispersion relations, crossing
symmetry, and analyticity. The bounds take the form of integrals over
the unphysical region $q^2>(M+m)^2$, which are then related to the
individual form factors in the physical region $0\leq q^2\leq \qmax$.
In Sec.\ 3 we derive our parametrizations by constructing quantities
from each form factor that can be legitimately expressed as Taylor
series with bounded coefficients. These parametrizations constitute
our main results.  Some technical issues are addressed in Sec.\ 4,
where we demonstrate that the error incurred by ignoring cuts in the
form factors is negligible.  Section 5 enumerates the corrections
to the parametrization and estimates their effects on the bounds.  We
discuss the sensitivity of our method to such corrections and
demonstrate that their effects are minimal, thus establishing the
robustness of the technique.  In Sec.\ 6 we use heavy quark symmetry
to relate the separate form factors appearing in the measured rate,
but point out that measurements of individual form factors in the near
future will obviate the need for this use of heavy quark symmetries.
We present results from fits of current data to this parametrization,
including values for $|V_{cb}| \CF (1)$, in Sec.\ 7. Our conclusions
appear in Sec.\ 8.

\newsec{Dispersion Relations}
The QCD matrix elements governing the semileptonic decays $\bar B \to
D^* l \bar \nu $ and $\bar B \to D l \nu$ may be expressed in terms of
the form factors
\eqn\fdefs{\eqalign{
  \bra{D^*(p',\eps)} V^\mu \ket{\bar B(p)} =&
    \, i g \epsilon^{\mu \alpha \beta \gamma} \epsilon_\alpha^* \, 
                          p_{\beta}' \, p_{\gamma} \cr
  \bra{D^*(p',\eps)} A^\mu \ket{\bar B(p)} =&
   \, f_0 \epsilon^{*\mu} + (\epsilon^{*} \! \cdot p) 
   [a_+ (p + p')^\mu + a_- (p - p')^\mu] \cr
   \bra{D(p')} V^\mu \ket{\bar B(p)} =& \, f_+ (p + p')^\mu + f_-
   (p - p')^\mu \cr }}
where $V^\mu = \bar c \gamma^\mu b$, and $A^\mu = \bar c \gamma^\mu
\gamma_5 b$. 
In terms of these form factors, the differential decay widths for
$\bar B \to D l \bar \nu$ and $\bar B \to D^* l \bar \nu$ are
respectively
\eqn\rated{
{d\Gamma \over dq^2}
 = {|V_{cb}|^2 G_F^2 (k^2 q^2)^{\frac32} \over 24 \pi^3 M^3} |f_+|^2
}
and
\eqn\rate{
{d\Gamma \over dq^2} = {|V_{cb}|^2 G_F^2 \sqrt{k^2 q^2} 
            \over 96 \pi^3 M^3} \bigl[2 q^2  |f_0|^2 
      + |F_1|^2 + 2 q^4 k^2 |g|^2\bigr],
}
where
\eqn\fcdef{
F_1= {1\over m} \left[2q^2 k^2 a_+ - \frac12 (q^2-M^2+m^2)f_0
\right] }
determines the partial width to longitudinally polarized $D^*$'s, and
$f_0$ and $g$ respectively determine the axial and vector
contributions from transversely polarized $D^*$'s (longitudinal
polarizations do not contribute to the vector matrix element in the
$\bar B$ rest frame, as is readily seen from Eq.~\fdefs).  $k^2$ is
related to the three-momentum squared ${\bf p}_D^2$ for $D$ or $D^*$
in the $\bar B$ rest frame, and is given by
\eqn\kdef{
k^2 = {M^2 \over q^2} {\bf p}_D^2 = {1\over 4 q^2}[q^2 - (M + m)^2]
[q^2 - (M - m)^2], }
with $M$ and $m$ the $\bar B$ and $D$ or $D^*$ meson masses,
respectively.

In our derivation of constraints from dispersion relations, we follow
the well-known methods developed by authors listed in Ref.~\hist.  We
begin by considering the two-point function
\eqn\twopntfnctn{
\Pi^{\mu \nu}_J (q) = (q^\mu q^\nu-q^2g^{\mu\nu})\Pi_J^T(q^2) +
g^{\mu\nu}\Pi_J^L(q^2) \equiv {i \int d^4\!x \, e^{iqx}\vev{0|{\rm T}
J^\mu(x) J^{\dagger\nu}(0)|0}, }}
where $J = V$ or $A$.  In QCD we can render both sides of this
relation finite by making one subtraction.  We thus obtain the
once-subtracted dispersion relations
\eqn\disper{
\chi^{T,L}_J (q^2)\equiv\left.{{\partial\Pi_J^{T,L}}
\over{\partial q^2}}\right.=
{1\over\pi}\int_0^\infty dt \, {{{\rm
Im}\,\Pi_J^{T,L}(t)}\over{(t-q^2)^2}}
.}
The functions $\chi^{T,L}_J(q^2)$ may be computed reliably in
perturbative QCD for values of $q^2$ far from the kinematic region
where the current $J$ can create resonances: specifically,
$(m_b+m_c)\Lambda_{\rm QCD} \ll (m_b+m_c)^2 - q^2$.  For resonances
containing a heavy quark, it is sufficient to take $q^2=0$.

The absorptive part ${\rm Im}\,\Pi_J^{\mu \nu}(q^2)$ is obtained by
inserting on-shell states between the two currents on the right-hand
side of Eq.~\twopntfnctn.  For $\mu = \nu$, this is a sum of
positive-definite terms, so one can obtain strict inequalities by
concentrating on the term with intermediate states of $B$-$D$ or
$B$-$D^*$ pairs.  The contribution of $B$-$D^*$ pairs to the
right-hand side of \disper\ enters (no sum on $\mu$) as
\eqn\optical{\eqalign{
{\rm Im}\,\Pi^{\mu \mu}_J (t=q^2) \geq & \, \frac{n_f}{2} \int d\Omega
       {\sqrt{k^2} \over 16 \pi^2 \sqrt{q^2}} \, \theta(q^2-(M+m)^2)
       \cr & \sum_{\epsilon} \vev{0| J^{\dagger \mu}|B({q\over2}-k)
       D^*({q\over2}+k,\eps)} \vev{B({q\over2}-k)
       D^*({q\over2}+k,\eps)| J^\mu|0}, }}
with an analogous form (no sum over polarizations) for $B$-$D$ pairs.
Here $n_f$ is the number of light valence quark flavors for the $B$
and $D$ or $D^*$ that give physically equivalent contributions; in
practice, we take $n_f=2$.  The momentum $q$ here and subsequently is
not to be confused with $q$ in Eq.~\disper, which will subsequently be
set to zero.  The matrix elements in Eq.~\optical\ are related by
crossing symmetry to those in Eq.~\fdefs.  That is, they are described
by the same form factors, but defined in different regions of the
complex $q^2$ plane.  $k^2$ is still defined by Eq.~\kdef\ but may now
be interpreted as the three-momentum squared of either the $B$ or
$D,D^*$ in the center of mass frame.  For massless leptons it turns
out that the partial widths appearing in Eq.~\rate\ present the same
combinations of form factors as the space-space components of
Eq.~\optical. It therefore suffices to use the dispersion relation
\eqn\drelation{
\chi^{\vphantom{\dagger}}_J
 = {1\over\pi}\int_0^\infty dt \, {{\rm Im}\,\Pi^{ii}_J(t)
\over{t^3}},
}
where $\chi^{\vphantom{\dagger}}_J = \chi^T_J(0) -
\frac12 \frac\partial{\partial q^2} \chi^L_J(0)$.  This definition of
$\chi^{\vphantom{\dagger}}_J$ corresponds to the combination of
$\Pi^T_J$ and $\Pi^L_J$ that gives $\Pi^{ii}_J$ at $q^2=0$.  At one
loop,
\eqn\pqcdchi{\eqalign{
\chi^{\vphantom{\dagger}}_V(u) = \chi^{\vphantom{\dagger}}_A(-u)
   = & \, {1 \over 32 \pi^2 m_b^2 (1 - u^2)^5 } \cr & \times [ (1 -
   u^2)(3+4 u- 21 u^2 + 40 u^3 - 21 u^4 + 4 u^5 + 3 u^6) \cr & \; + 12
   u^3 (2 - 3 u + 2 u^2)\ln u^2 ], \cr }}
where $u = { m_c \over m_b} $ is the ratio of quark masses.  For $u =
0.33$, $\chi^{\vphantom{\dagger}}_V = 9.6 \cdot 10^{-3}/m_b^2$ and
$\chi^{\vphantom{\dagger}}_A = 5.7 \cdot 10^{-3}/m_b^2$.

When substituted into Eq.~\drelation, \optical\ and \pqcdchi\ lead to
bounds on integrals of the analytically continued form factors.  For
example, for the axial current $J=A$, Eq.~\optical\ becomes
\eqn\twofs{
{\rm Im}\,\Pi^{ii}_A \geq {n_f \sqrt{k^2} \over 12 \pi \sqrt{q^2}}
\left[ |f_0|^2 + \frac1{2 q^2} |F_1|^2  \right]
\theta(q^2-(M+m)^2).
}
The bound in this case, which may be taken to constrain $|f_0|$ and
$|F_1|$ separately, reads
\eqn\egbd{
{n_f \over 12 \pi^2 \chi^{\vphantom{\dagger}}_A}
\int_{(M+m)^2}^{\infty} dq^2 \, \frac{\sqrt{k^2}}{(q^2)^\frac72}
\left[ |f_0|^2 + \frac{1}{2 q^2} |F_1|^2 \right] \leq 1 .
}

We now define a new variable $z$ by
\eqn\zdef{
{1+z \over 1-z} = \sqrt{(M+m)^2 -q^2 \over 4 M m}.
}
Taking the principal branch of the square root in this expression, the
change of variables $q^2 \to z$ maps the two sides of the cut $q^2 >
(M+m)^2 $ to the unit circle $|z|=1$, with the rest of the $q^2$ plane
mapped to the interior of the unit circle.  In particular, the real
values $-\infty < q^2 \leq (M-m)^2$ and $(M-m)^2 \leq q^2 < (M+m)^2$
are mapped to the real axis, $1 > z \geq 0$ and $0 \geq z > -1$
respectively.  Written in terms of $z$, the inequalities from
Eqs.~\optical\--\pqcdchi\ now read
\eqn\zbound{
{1\over2\pi i} \int_C {dz \over z} |\phi_i(z) F_i(z) |^2 \leq 1.
}
The contour $C$ is the unit circle.  The weighing functions are
\eqn\phij{
\phi_i = M^{2-s}2^{2+p} \sqrt{\kappa n_f }
        [r(1+z)]^{p+1 \over 2} (1-z)^{s - \frac32 }
     [(1-z)(1+r) + 2 \sqrt{r} (1+z)]^{-s-p},
}
where $r = m/M$ is the ratio of meson masses, and $\kappa$, $p$ and
$s$ depend on the form factors $F_i$ as listed in Table~1.
\bigskip
\vbox{\medskip
\hfil\vbox{\offinterlineskip
\hrule
\halign{&\vrule#&\strut\quad\hfil$#$\quad\cr
height2pt&\omit&&\omit&&\omit&&\omit&&\omit&\cr
&i &&F_i&& 1/\kappa\qquad && p && s& \cr
\noalign{\hrule}
&0&& f_0 && 12 \pi M^2\chi^{\vphantom{\dagger}}_A
&& 1 && 3 &\cr
&1 && F_1
&& 24 \pi M^2 \chi^{\vphantom{\dagger}}_A &&
1 &&4&\cr
&2&& g && 12 \pi M^2\chi^{\vphantom{\dagger}}_V
&& 3 &&1 &\cr
&3 && f_+ && 6 \pi M^2\chi^{\vphantom{\dagger}}_V
&& 3 &&2 &\cr }
\hrule}
\hfil}
\medskip
\INSERTCAP{1}{Factors entering Eq.~\phij\ for the form factors
$F_i$.}
The results \zbound\--\phij\ apply equally well to analogous
heavy-to-light form factors such as in $\bar B \to K^* \gamma$ and
$\bar B \to \pi l \nu$; for the latter process, they agree with
Ref.~\dec\ upon substitution of $m_\pi$ for $m_D$.

\newsec{Parametrization of Form Factors}
Our parametrizations of the form factors rely on a Taylor expansion
about $z=0$. To connect this expansion to bounds at $|z| =1$, we need
a function which is analytic inside the unit disk.  The form factors
$F_i$ have cuts and poles along the segment $q^2>(M-m)^2$ of the real
axis in the complex $q^2$ plane, and therefore only on the segment
$(-1,0)$ of the real axis in $z$ or on the unit circle $|z| = 1$.

We have used the freedom to redefine $\phi_i$ by a phase to ensure
that it has no poles, branch cuts, or zeros in the interior of the
unit circle $|z|<1$, but the form factors $F_i(q^2)$ have poles due to
the existence of stable spin-one states with unit bottom and charm
number (spin-zero states only contribute to $f_-$ and $a_-$, which,
for massless leptons, give vanishing contribution to the differential
rate). The masses of these $B_c^*$ mesons can be reliably
computed\refs{\klt{--}\ck}\ with potential models. The vector states
are predicted to have masses corresponding (for $z$ defined with $m =
m_{D^*}$) to $ z_1 = -0.284$, $z_2 = -0.472$, $z_3 = -0.531$, and $z_4
= -0.907$, while the axial vector masses correspond to $z_5 = -0.395$,
$z_6 = -0.399$, $z_7 = -0.609$, and $z_8 = -0.619$.  One may form
functions $P(z)$ that are products of terms of the form
$(z-z_i)/(1-\bar z_i z)$, known to mathematicians as Blaschke
factors\blasref:
\eqn\blaschke{\eqalign{
 P_0 &= P_1 =  \prod_{j=5}^8 {(z -z_j) \over
    (1 - \bar z_j z)} , \cr
 P_2 &= P_3  =  \prod_{j=1}^4 {(z -z_j) \over
    (1 - \bar z_j z)}.\cr
}}
Such $P_i$'s are analytic on the unit disk for $|z_j|<1$ and serve to
eliminate poles of $F_i$ at each $z=z_j$ when formed into the products
$P_i(z) F_i(z)$.  Most importantly, each $P_i$ is unimodular on the
unit circle, and therefore we may replace $F_i$ with $P_i F_i$ in our
bound Eq.~\zbound\ without changing the result.  Since now both $P_i
F_i$ and $\phi_i$ are analytic on the unit disc, Taylor expanding
$\phi_i P_i F_i$ about $z=0$ gives
\eqn\master{
F_i(z) = {1\over P_i(z) \phi_i(z)} \sum_{n=0}^\infty a_n z^n . 
}
Substituting this expression into Eq.~\zbound\ gives the central
result
\eqn\asum{
\sum_{n=0}^\infty |a_n|^2 \leq 1.
}
The coefficients $a_n$ are different for each form factor, and must be
determined by experiment.  However, since both $B$-$D^*$ and $B$-$D$
states contribute to the same vector-vector dispersion relation, the
sum of the squared-coefficient sums for $f_+$ and $g$ is bounded by
one:
\eqn\fg{
\sum_{n=0}^\infty (|a^{(f_+)}_n|^2 + |a^{(g)}_n|^2) \leq 1.
}
This relation holds if $z$ in Eq.~\master\ is defined using $m =
m_{D^*} $ for $g$ and $m = m_D$ for $f_+$.  An analogous result
constrains the coefficients $a_n$ of the form factors $f_0$ and $F_1$,
\eqn\ff{
\sum_{n=0}^\infty (|a^{(f_0)}_n|^2 + |a^{(F_1)}_n|^2) \leq 1.
}
For the remainder of this paper, we content ourselves with the weaker
constraint Eq.~\asum .

\def\zmax{{z_{\rm max}}}

The utility of this parametrization arises from the
observation that the physical range $q^2_{\rm max} \geq q^2
\geq 0$ for $\bar B \to D^* l
\bar \nu$ $(D l \bar \nu)$
semileptonic decays corresponds to $0
\leq z \leq \zmax=0.056 (0.065)$.  We
define an approximation $F_i^N$ to
the form factor $F_i$ by truncating
after the $N$th term:
\eqn\trunc{
F_i^N (z) = \frac{1}{P_i(z) \phi_i(z)} \sum^N_{n=0} a_n z^n .
}
Then the maximum error incurred by truncating after $N$ terms is just
\eqn\err{\eqalign{
{\rm max} |F_i(z)-F_i^N(z)| & = \frac{1}{|P_i(z) \phi_i(z)|}
\sum^{\infty}_{n=N+1} \! |a_n| \, z^n \cr 
& \leq \frac{1}
{|P_i(z) \phi_i(z)|} \sqrt{\sum_{n=N+1}^{\infty} \! |a_n|^2}
\sqrt{\sum_{n=N+1}^{\infty} \! z^{2n}} \cr 
& < \frac{1}{|P_i(\zmax) \phi_i(\zmax)|}
\frac{z_{\rm max}^{N+1}}{\sqrt{1-z_{\rm max}^2}},  
}}
where we have used the Schwarz inequality, Eq.~\asum, and the fact
that $z^{N+1}/|P_i(z)\phi_i(z)|$ increases
monotonically over the physical range.  For $N$ as small as 2, this
truncation error is quite small; see Table~2.

To calculate a relative error we need to estimate the form factor
itself. This can be done at $z=0$ using heavy quark symmetries. The
resulting bound on the relative error, $ |F_i(z)-F_i^N(z)|/F_i(0)$, is
shown in Table~2.

The larger relative error associated with $F_1$ arises from a
collusion of factors. Compared to $f_0$, these consist of a smaller
value of $\kappa$ and a greater value of $s+p$, both of which decrease
$\phi_1$, as well as a smaller normalization $F_1(0)$. The accuracy of
the parametrization of $F_1$ is improved to $0.34\%$ by truncating
after one more parameter (\ie, taking $ N=3$ above).

\bigskip
\vbox{\medskip
\hfil\vbox{\offinterlineskip
\hrule
\halign{&\vrule#&\strut\quad\hfil$#$\hfil\quad\cr
height2pt&\omit&&\omit&&\omit&&\omit&&\omit&\cr
&i &&F_i&& |F_i(z)-F_i^N(z)|\times10^2 
&& |F_i(z)-F_i^N(z)|/F_i(0) & \cr
\noalign{\hrule}
&0&& f_0 && 1.2 && 1.0\% &\cr
&1 && F_1 && 4.6 &&6.1\%&\cr
&2&& g && 0.8  &&0.5\% &\cr
&3 && f_+ && 1.4 &&1.3\% &\cr }
\hrule}
\hfil}
\medskip
\INSERTCAP{2}{Bounds on truncation errors, $ |F_i(z)-F_i^N(z)|$, for
$N=2$, for various form factors from Eq.~\err. To estimate a
corresponding relative error, we use the value of the form factor at
threshold, $F_i(0)$, as predicted by heavy quark symmetries.}

\newsec{Branch Cuts}
In the previous section we ignored branch cuts in the form factors
with branch points inside $|z|=1$. These cuts originate from
non-resonant contributions with invariant masses below $M+m$. For
example, branch points are expected at $q^2=(m_{B^*_c}+n \, m_\pi)^2$,
with $n$ a positive integer, and at $q^2=(m_{\eta_{bc}}+m_\rho)^2$,
where $\eta_{bc}$ is the pseudoscalar partner of the vector
$B_c^*$. We now show that their neglect is quite justified.

We content ourselves with estimating the effect of any single cut
modeled in a reasonable way, since multiple cuts can be handled
analogously, and cuts modeled differently give comparable
results.\footnote{$^\dagger$}{The statement in Ref.~\bglapr\ that the
effect of such cuts may be incorporated by mapping them onto the unit
circle and expanding in a new basis is erroneous; the matching of
coefficients in the new basis to a Taylor expansion about $z=0$
involves an infinite number of equally important terms.}

Any form factor $g(q^2)$ in Eq.~\fdefs\ satisfies a simple dispersion
relation
\eqn\gdisp{
g(q^2)={1\over\pi}\int_0^\infty {\rm d}t\,{{\rm Im }\,g(t)\over
t-q^2}.
}
A reasonable model for a cut can be obtained, say, by taking an
additive contribution to $g$ satisfying
\eqn\cutform{
{\rm Im} \, g(t) = C \left( \sqrt{t-M_b^2} \,\theta(t-M_b^2)
-\sqrt{t-M_{\vphantom{\dagger}a}^2} \, \theta(t-M_a^2) \right),
}
where $q^2 = M_a^2$ is the location of the branch point of interest,
and $M_b$ is an arbitrary scale with $M_b^2 > M_a^2$.  The subtraction
is performed to ensure that ${\rm Im }\,g(t)$ vanishes as
$t\to\infty$. The precise form of the subtraction is immaterial, so we
choose one that simplifies our calculations. Moreover, since branch
points on $|z|=1$ are irrelevant, we need consider only the case $M_b<
M+m$. In all cases $M_a>m_{B_c^*}+m_\pi$, the location of the lowest
branch point, which has $z = -0.32$ for $m=m_{D^*}$.

The coefficient $C$ arises as a coupling in diagrams connecting the
$(V-A)$ current to an external $B$-$D$ or $B$-$D^*$ pair through
non-resonant on-shell intermediate states.
\OMIT{
By perturbative unitarity, such contributions are given by the
imaginary part of the amplitude and may be evaluated by ``cutting''
the diagram to obtain one in which the intermediate states become
asymptotic states.  Using this reasoning, let us estimate the size of
$c$ for a two-particle cut.  } The intermediate states couple to the
current with a strength $\hat{f}$, and to $B$-$D$ or $B$-$D^*$ with
strength $\hat{g}$.  Furthermore, two-particle phase space provides a
factor of $1/8\pi$.  Phenomenologically $C \approx \hat{f}\hat{g} /
8\pi$ is expected to be quite small.  We consider the most extreme
case, namely, $C=M^{s-3} c$ where $c$ is dimensionless and at most of
order unity.

Writing our model cut from Eq.~\cutform\ in terms of the variable $z$,
we have
\eqn\cutforminz{
g_{\rm cut}(z)= 4cM^{s-2} \sqrt{r} \left({\sqrt{(z-z_a)(1-zz_a)}\over
(1-z)(1-z_a)} - {\sqrt{(z-z_b)(1-zz_b)}\over (1-z)(1-z_b)} \right). }
Let $f(z)$ stand for any of the form factors, with corresponding
functions $\phi(z)$ from Eq.~\phij\ and $P(z)$ from
Eq.~\blaschke. Consider the difference $\tilde f(z) = f(z) - g_{\rm
cut}(z)$, and let $f_{\rm cut} = g_{\rm cut} \phi P$.  The function
$\tilde f \phi P$ is thus designed to be analytic on the unit disc.
We proceed in two steps: First we find a bound for $\tilde f \phi$
analogous to that in \zbound; this constraint translates into a new
bound on the parameters in our expansion.  Then we show that $f_{\rm
cut}$ is well approximated in the physical region by a polynomial of
low degree, so that truncating our expansion after only a few terms
incurs a very small error.

By the Minkowski inequality and~\zbound\ we have
\eqn\minkowski{
\left(\int_0^{2\pi} d\theta\, |\tilde f \phi|^2\right)^{1/2}\le
\left(\int_0^{2\pi} d\theta\, |f \phi|^2\right)^{1/2}+
\left(\int_0^{2\pi} d\theta\, |f_{\rm cut}|^2\right)^{1/2} \le
\sqrt{2\pi}(1 + I_{\rm cut}^{1/2}), }
where $z=e^{i\theta}$ and
\eqn\Idefd{
I_{\rm cut}\equiv{1\over{2\pi}}\int_0^{2\pi} d\theta\, |f_{\rm cut}|^2~.
}
As before, the functions $P(z)$ are unimodular on the unit circle, and
so leave the integrals unchanged.  $I_{\rm cut}$ can be computed using
the explicit form for the cut in Eq.~\cutforminz.  As a specific
example, take for $\phi$ in Eq.~\phij\ the case of $p=3$ and $s=1$,
corresponding to the form factor $g$.  The numbers to follow are
specific to the case $m = m_{D^*}$, although the qualitative results
are the same for $m=m_D$.  For $z_a=-0.32$ and $z_b=-1.0$, we find
\eqn\Icutresult{
I_{\rm cut}
\approx 2.2 \times10^{-3} c^2.
}
Thus the bound on $\tilde f$ is relaxed only by $\sim5\%$ times $c$
relative to that on $f$. A realistic choice of coefficient $c$
significantly improves this bound, as does a branch point closer to
the $B$-$D$ or $B$-$D^*$ threshold $z=-1$.  For example, for
$z_a=-0.5$ replace $2.2\times10^{-3}$ by $4.0\times10^{-4}$ in
Eq.~\Icutresult.

Being analytic, $\tilde f \phi P$ has a Taylor expansion for
$|z|\le1$. Hence one may write
\eqn\newexpansion{
f(z)={1 \over \phi(z) P(z)} \left( \sum_{n=0}^\infty a_nz^n + f_{\rm
cut} \right).
}
Here the coefficients $a_n$ are bounded, $\sum_{n=0}^\infty
|a_n|^2\le(1+I_{\rm cut}^{1/2})^2$. Moreover, $f_{\rm cut}$ is
analytic over the physical region. Define the remainder $R^{N}_{\rm
cut}$ through
\eqn\fcutexpand{
f_{\rm cut}(z) = \sum_{n=0}^\infty b_n z^n = \sum_{n=0}^{N} b_n z^n
+ R^{N}_{\rm cut}(z)~.  }
The remainder $R^{N}_{\rm cut}$ is the additional error introduced in
the parametrization of $f(z)$ as an $N$-th order polynomial in $z$
with coefficients $a_n+b_n$. These coefficients obey a constraint
almost identical to Eq.~\asum, because the $b_n/c$ are uniformly
small: $\sum_{n=0}^{n=N}|a_n+b_n|^2\le (1+I_{\rm
cut}^{1/2})^2(1+\sum_{n=0}^{n=N}|b_n|)^2= (1.07)^2$, $ (1.07)^2$, and 
$(1.08)^2$, for $c=1$ and $N=2$, 3, and 4, respectively.  The maximum
of $R^{N}_{\rm cut}(z)$ over the physical region is $-1\times10^{-6}$,
$2\times10^{-7}$, and $4\times10^{-9}$ times $c$ for $N=2$, 3, and~4,
respectively. These figures should be compared with the bound on the
truncation in the analytic part, $\sum_{n=N+1}^{\infty} a_n z^n \leq
(1+I_{\rm cut}^{1/2}) (0.056)^{N+1}\approx 2\times10^{-4}$,
$1\times10^{-5}$, and $6\times10^{-7}$. Since we expect realistic
branch cuts to have $c\ll1$, we see that their effect is negligible.

\newsec{Uncertainties}
 
Two important statements follow immediately from Eqs.~\master\
and~\asum :
\item{a~)}   Each of the various $\bar B \to D$ or $\bar B \to D^*$
form factors can be accurately described by three parameters, one of
which is the normalization at zero recoil, with a truncation error of
order 1\%;
\item{b~)}   The fitting parameters obey $ \Sigma_{n=0}^{n=2}
|a_n|^2 \leq B^2$, with $B =1$.

A number of approximations have been made in deriving these results.
How do corrections to these approximations alter the above statements?
We answer this question by noting that nearly all the corrections we
expect to be non-negligible can be taken into account by altering the
bound to $B \ne 1$.

An estimate of how much $B$ might change may be made by considering
uncertainties arising from the following sources:

\item{1~)} The $b$ and $c$ quark masses, which enter into the
one-loop perturbative functions $\chi^{\vphantom{\dagger}}_J$, are not
well established; we take $m_c/m_b=0.33$. This leads to a roughly $5
\%$ uncertainty in the normalization of $\phi$ which, by redefining
the $a_n$, is equivalent to a $5 \%$ uncertainty in the value $B$
bounding the $a_n$.

\item{2~)} The functions $\chi^{\vphantom{\dagger}}_J$ 
also receive perturbative two-loop corrections. Since $\phi$, and
therefore the bound, depends only on
$(\chi^{\vphantom{\dagger}}_J)^{1/2}$, such corrections should lead to
no more than a $15 \%$ change in $B$.

\item{3~)} The masses of the $B_c^*$ poles were computed from
a potential model.  Computations by various groups typically agree to
a fraction of a percent\refs{\klt{--}\ck}.  The results from two
different groups\refs{\klt{,}\eq}\ give Blaschke factors that agree to
$2\%$. However, $P(z)$ is sensitive to the mass of the $3S$ vector
state, which is close to the $B$-$D^*$ threshold, and is presented
only by \eq. Changing it by $1\%$ results in a $20\% $ change in
$P(0)$; $P(z)/P(0)$, however, varies by less than $1 \%$.
 
\item{4~)}  We argued in the previous section that contributions
from multi-particle cuts should alter the bound $B$ by less than
$8\%$.

\item{5~)}  In extracting values of $|V_{cb}|\CF(1) $, we require
bounds not on $a_1$ and $a_2$ alone, but on $a_1 / \CF(1)$ and $a_2 /
\CF(1)$, and these depend on the zero-recoil normalization $\CF(1)$.
This normalization is predicted to no worse than 20\% accuracy by
heavy quark symmetry\thresh.

\smallskip
The uncertainties (1) to (5) are uncorrelated, so to estimate their
total effect, we add them in quadrature. This leads to a relaxation on
our bounds from $B=1$ to less than $B = 1.4$.

This relaxation of the bound increases the truncation error on, for
example, $f_0$ from $0.012$ to $0.017$, still negligibly small given
the current experimental accuracy. Even if we added the uncertainties
linearly, the truncation error would only rise to $0.020$.  We see
that statement (a) is extraordinarily robust; it is nearly independent
of the size of the uncertainties listed above.

On the other hand, allowing a value of $B$ larger than 1 in statement
(b) could in principle affect the extraction of $|V_{cb}|$
significantly due to a larger allowed range for $a_1$ and $a_2$.  For
this reason, it would be useful to pin down the $\alpha_s$ corrections
and the mass of the $3 S$ vector $B_c^*$ state more precisely.
However, for the extraction we perform in Sec.\ 7, relaxing the bound
from $B=1$ to 1.4 turns out to change the results very slightly: The
central values of $|V_{cb}|\CF(1)$ and the slope change by no more
than a tenth of a standard deviation.

One should also bear in mind that our bounds can be significantly
improved by the inclusion of more terms than just $B$-$D^*$ or $B$-$D$
pairs to saturate the bound in Eq.~\optical.  Such contributions arise
through higher resonances of the current $J$; if estimated
numerically, they have an effect equivalent to reducing $B$.

\newsec{Heavy Quark Symmetry}

The parametrizations Eq.~\master\ make no use of heavy quark symmetry.
Thus, $1/m_c$ corrections to the extraction of $|V_{cb}|$ from $\bar B
\to D l \bar \nu$ decays enter only through the normalization of the
form factor $f_+(z=0)$ at zero recoil. This normalization is
determined by heavy quark symmetry to $\CO (1/m_c)$.

If the individual $\bar B \to D^* l \bar \nu$ form factors $f_0$,
$F_1,$ and $g$ are experimentally determined in the near future,
separate extractions of $|V_{cb}|$ can be made for each form factor.
These extractions will depend on heavy quark symmetry only through the
normalization of form factors at zero recoil.  This is useful because
the normalization of $f_0$ is predicted to $\CO (1/m_c^2)$\luke.

At present, to extract $|V_{cb}|$ from the $\bar B \to D^* l \bar \nu$
differential width~\rate\ in terms of our three-parameter
descriptions, one must relate $f_0$, $a_+$, and $g$ using heavy quark
symmetry.  In the infinite mass limit, all form factors for $\bar B
\to D$ and $\bar B \to D^*$ (as well as $B \to B$) are directly
proportional to the universal Isgur-Wise function\iw.  Consequently,
the ratio of any two form factors assumes a simple form:
\eqn\ffrat{
a_+ /g = -\frac12, \qquad f_0/a_+ = -2 M^2 r (\w + 1), \qquad
{\rm and}
\quad f_0 / g = M^2 r (\w +1), }
where $\w = v \cdot v'$ is the product of the $\bar B$ with $D$ or
$D^*$ meson velocities.  These ratios admit two types of correction,
namely power corrections in $1/m_c$, and running and matching
corrections relating QCD to the heavy quark effective theory.  We
discuss each of these in turn.

Because many heavy quark symmetry-violating contributions cancel in
the above ratios, one might expect $1/m_c$ corrections to be smaller
than in, say, the relation between the $B \to B$ elastic and $\bar B
\to D$ or $\bar B \to D^*$ semileptonic form factors.  For example, 
$a_+ /g = -\frac12$ may be derived using only charm quark spin
symmetry, without recourse to bottom-charm flavor symmetry; spin
symmetry is expected to hold more precisely than the full flavor-spin
symmetry\hqben.  In addition, the ratio $f_0/ g = M^2 r (\w +1)[1 +
(\w-2) \bar \Lambda / 2m_c ]$ involves no unknown $\w$-dependent
functions\luke\ at $\CO(1/m_c)$, but only the constant $\bar \Lambda =
m_D - m_c$. The third ratio is given by the quotient of these two.
Choosing two different pairs of the above ratios gives two different
parametrizations of the decay form factor $\CF(\w)$ conventionally
defined by
\eqn\cfdef{
{d\Gamma \over d\w} = {|V_{cb}|^2 G_F^2 \over 48 \pi^3} m^3 (M-m)^2
         \sqrt{\w^2 -1}[4 \w (\w+1) {1 - 2 \w r + r^2 \over (1-r)^2} +
         (\w+1)^2 ] \CF^2(\w) .
}
In the heavy quark limit, the form factor $\CF(\omega)$ is simply the
Isgur-Wise function times QCD corrections (discussed below), and we
readily see that Eq.~\rate\ reduces to Eq.~\cfdef. In terms of the
parametrizations Eq.~\master\ of $g$ and $f_0$, respectively,
\eqna\scriptf
$$\eqalignno{
\CF(z)&={1 \over P_2(z) \phi_2(z) }\bigl[ P_2(0) \phi_2(0) \CF(z=0)
  + M\sqrt{r}(a_1 z + a_2 z^2 )\bigr] &\scriptf a\cr 
&={(1-z)^2 \over P_0(z) \phi_0(z) (1+z)^2 }
\bigl[ P_0(0) \phi_0(0) \CF(z=0) +{1\over2M\sqrt{r}}
(b_1 z + b_2 z^2 )\bigr]{}. &\scriptf b\cr } 
$$
The form factor may be expressed as a function $\CF(\w)$ of velocity
transfer by rewriting $z$ as
\eqn\zdefw{
z = { \sqrt{\omega + 1 \over 2} -1 \over \sqrt{ \omega + 1 \over 2}
               +1}.
}
At zero recoil, $\CF(\w=1) =1$ times corrections whose estimates range
from 0.89 to 0.99\thresh. Relative to this normalization, the 
parametrization in Eq.~\scriptf a\ has a $0.5\%$
truncation error, while that in  Eq.~\scriptf b\
has a $1.0\%$ truncation error; see Table~2.

To the degree that $1/m_c$ corrections are negligible, the
extracted values of $|V_{cb}|\CF(\w = \! \! 1 \! )$ and the
slope $\CF '(\w =1)$ cannot depend on which of the
parametrizations \scriptf\null\ we use. Since $1/m_c$
corrections enter differently into each of these
parametrizations, the degree to which this is true gauges the
sensitivity of the extraction to heavy quark symmetry
violations.

For a thorough accounting of relations between form factors when using
heavy quark symmetry, one must also include effects due to the running
of the QCD coupling $\alpha_s$ and matching between the full theory of
QCD and the heavy quark effective theory.  The form factors are then
no longer just trivial factors times the Isgur-Wise function, but now
include a functional dependence on $\omega$, as well as $m_c$, $m_b$,
and the value of $\alpha_s$ at these scales.  For conciseness and
definiteness, we adopt the notation of Neubert\neubrev\ to parametrize
such corrections.  The relation between the Isgur-Wise function
$\xi(\omega)$ and the relevant form factors then reads
\eqn\match{
a_+ = -\frac{1}{2M\sqrt{r}} (\hat{C}^5_1 + \hat{C}^5_2 r +
\hat{C}^5_3) \xi, \qquad f_0 = M \sqrt{r} (\omega + 1) \hat{C}_1^5 \xi,
\qquad g = \frac{1}{M\sqrt{r}} \hat{C}_1 \xi .
}
The functions $\hat{C}_1$, $\hat{C}^5_1$ become unity when the strong
coupling is switched off, whereas the other $\hat{C}$'s vanish.  In
this limit we recover the ratios in Eq.~\ffrat.

Apart from changing the overall normalization of form factors at zero
recoil by a few percent, the functional dependences in Eq.~\match\
turn out to be rather weak over the allowed range for $\bar B \to D$
or $\bar B\to D^*$ semileptonic decay ($\omega = 1.0$ to $1.5$).  In
particular, $\hat{C}_1$ decreases from 1.136 to 1.011 over this range,
but $-2a_+/g = 0.864 \to 0.882$, and $f_0/g M^2 r (\omega+1) = 0.868
\to 0.884$ are nearly constant.  In addition, corrections due to
running between the bottom and charm mass scales cancel out of such
ratios.

Because the undetermined $1/m_c$ corrections are just as significant,
there is little to be gained in incorporating the calculated matching
corrections explicitly in our analysis; rather, our sensitivity to
both $1/m_c$ and matching corrections is gauged by comparing the
extractions of $|V_{cb}|\CF(1)$ and the slope $\CF '(1)$ by the two
parametrizations of Eq.~\scriptf\null. Compared to the $g$ parametrization,
the $f_0$ parametrization changes the central values of CLEO's
$|V_{cb}| \CF(1)$, and both CLEO's and ARGUS's $\CF'(1)$ by less than
a fourth of a standard deviation; ARGUS's and ALEPH's $|V_{cb}|
\CF(1)$, as well as ALEPH's $\CF'(1)$, change by less than a tenth of
a standard deviation ({\it i.e.}, $<2\%$ for all $|V_{cb}|
\CF(1)$).

\newsec{Results}

Since both parametrizations \scriptf\null\ give essentially the same
results, we choose the $g$ parametrization, which has a smaller
truncation error. 
From the point of view of heavy quark symmetry, one should use the
$f_0$ parametrization, since $f_0(\w=1)$ is predicted to higher
accuracy. Here we are more concerned with exploring the implications
of our parametrizations.  The central values and $68 \%$ confidence
intervals should be taken as indicative; proper inclusion of
efficiencies, resolutions, and correlated errors can only be done by
the experimental groups themselves.

Fitting $|V_{cb}| \CF(1)$, $a_1/ \CF(1)$, and $a_2/ \CF(1)$ to
experiment yields the results in Table~3. For each experiment, we have
listed the best fit values for $|V_{cb}| \CF(1)$, $a_1/ \CF(1)$, and
$a_2/ \CF(1)$, as well as the resulting slope $\CF'(\w =1)$. The
$68\%$ confidence intervals due to statistics are included as well.
The parametrization \scriptf a\ includes the constraint
$\sum_{n=0}^\infty |a_n|^2 \leq 1$; for comparison, we also present
the best fit values resulting from an unconstrained fit with freely
varying $a_n$.
\bigskip
\vbox{\medskip
\hfil\vbox{\offinterlineskip
\hrule
\halign{&\vrule#&\strut\quad\hfil$#$\quad\cr
height2pt&\omit&&\omit&&\omit&&\omit&&\omit&&\omit&\cr
&B&&|V_{cb}|\CF(1)\cdot 10^3&&a_1/ \CF(1)\hfil&&a_2/
\CF(1)\hfil&&
\CF'(1) &&\rm{Expt.}\hfil&\cr
height3pt&\omit&&\omit&&\omit&&\omit&&\omit&&\omit&\cr
\noalign{\hrule}
height2pt&\omit&&\omit&&\omit&&\omit&&\omit&&\omit&\cr &  1
&&35.7_{-2.8}^{+3.7}&&0.046_{-0.14}^{+0.05}&&-1.00_{-0.0}^{+2.0}&&
-0.89^{+0.3}_{-0.8}&&\rm{CLEO} &\cr &  \infty
&&33.3_{-6.1}^{+6.1}&&0.181_{-0.27}^{+0.38}&&-3.20_{-5.9}^{+4.5}&& -
0.14^{+2.1}_{-1.5} &&\rm{CLEO} &\cr
\noalign{\hrule}
& 1 &&45.8_{-10.9}^{+8.1}&&-0.200_{-0.07}^{+0.22}
&&0.98_{-2.0}^{+0.0}&&-2.3^{+1.2}_{-0.4}&& \rm{ARGUS} &\cr
& \infty &&49.5_{-19.5}^{+19.4}&&-0.297_{-0.32}^{+0.71}
&&2.59_{-11.3}^{\,\,\,+5.4}&&-2.8^{+3.9}_{-1.8}&& \rm{ARGUS} &\cr
\noalign{\hrule}
& 1
&&31.5_{-5.8}^{+4.5}&&0.090_{-0.10}^{+0.25}&&1.00_{-2.0}^{+0.0}&&
-0.65^{+1.4}_{-0.6} &&\rm{ALEPH} &\cr & \infty
&&31.8_{-7.5}^{+7.5}&&0.073_{-0.33}^{+0.52}&&1.33_{-7.8}^{+5.3}&&
-0.74^{+2.9}_{-1.8} &&\rm{ALEPH} &\cr }
\hrule}
\hfil}
\bigskip
\INSERTCAP{3}{Fit values for $|V_{cb}| \CF(1)$, $a_1/ \CF(1)$, 
$a_2/ \CF(1)$, and the zero recoil slope of $\CF(\w)$ from the various
experiments, constrained to obey $\Sigma_{n=0}^{n=2} |a_n|^2 \leq B $.
}
The fits allowed by QCD are those with (in particular) $|a_2| \leq
1$. The extracted values of $|V_{cb}|$ are in good agreement with a
previous extraction\bglapr, after accounting for differences in
definitions and experimental data.  We have renormalized the ARGUS
data to bring their assumed $B$ lifetime and $D^0 \to K^- \pi^+$
branching ratio into agreement with more recent experiments; we use
$\tau_B = 1.61$ psec and\pdg\ $B(D^0 \to K^- \pi^+) = 4.01 \% $.

The central values for $|V_{cb}| \CF(1)$ agree surprisingly closely
with those of the experimental groups themselves. This did not need to
be the case, as one can see from the behavior of the unconstrained
fit.

The connection between our parameters $a_1,a_2$ and the commonly
used expansion in $(\w -1)$ is
\def\ao{ \frac{a_1}{\CF(1)} }
\def\at{ \frac{a_2}{\CF(1)} }
\eqn\wexpand{\eqalign{
{\CF(\w) \over \CF(1) } &= 1 + \left[5.54 \ao - 1.15\right] (\w -1) +
   \left[-7.73 \ao + 0.69 \at + 1.11\right](\w -1)^2 \cr &+ \left[8.19
   \ao - 1.14 \at - 0.99\right] (\w-1)^3 + ...  }
}
While such an expansion describes the form factor well close to zero
recoil, it converges poorly over the rest of the kinematic
range. Substituting the allowed range of parameters $\sum_{n=0}^{n=2}
|a_n|^2 \leq 1$ gives a truncation error for a quadratic fit in $(\w
-1)$ of $120 \%$; the truncation error of a linear fit is $220 \%$.
To be assured of fitting a QCD-allowed form factor at percent-level
accuracy, a parametrization obeying the same constraints as
Eq.~\scriptf\null\ must be used.

Plotted in Fig.\ 1 are the  constrained and unconstrained  fits to the
CLEO\cleo\ data.  Both fits match the data well; the chi-squares per
degree of freedom are $\chi^2 /{\it dof} = 0.65$ and $0.50$,
respectively.  The CLEO group extracts $|V_{cb}| \CF(1)
\cdot 10^3 = 35.1 \pm 1.9 \, ({\it stat})$ and a slope $\CF'(1) =
-0.84 \pm 0.13 $ using a linear fit, in close agreement with our
bounded fit. The unbounded fit serves as an illustration of a
parametrization which gives a markedly different best fit; the central
value of $|V_{cb}| \CF(1)$ differs by $5\%$ from the linear result,
while the slope is in violation of the Bjorken bound\bjbd, $\CF'(1) <
-1/4$.  By Eq.~\asum, this uncontrained  fit is ruled out by QCD.

\INSERTFIG{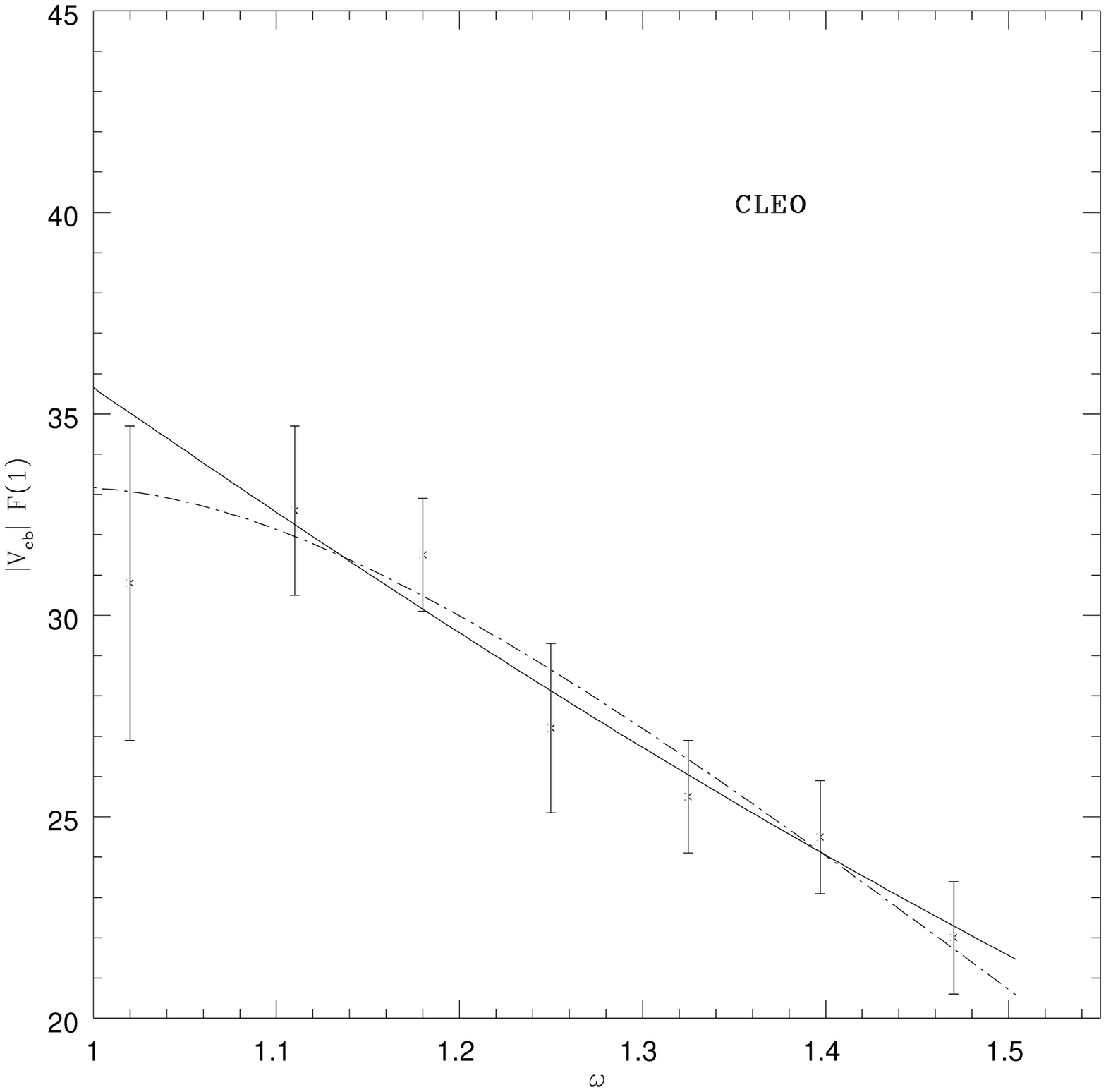}{1}{Fit of CLEO data to our parametrization,
Eq.~\scriptf a. The solid line shows the result of imposing the QCD-derived
constraint $\sum_{n=0}^{n=2} |a_n|^2\le1$ on the parametrization. The
dot-dash line shows the corresponding
unconstrained fit.}

For ALEPH\aleph\ the constrained and unconstrained fits overlay each other
quite closely (Fig.\ 2).  A linear fit by the ALEPH group gives
$|V_{cb}| \CF(1) \cdot 10^3 = 31.4 \pm 2.3 \, ({\it stat})$ and a
slope $\CF'(1) = -0.39 \pm 0.21 $, in good agreement with the results
of our constrained fit. The confidence intervals in Table~3 are
somewhat larger for ALEPH than might be expected because of the
smallness of the minimum $\chi^2$: Both the bounded and unbounded fits
have $\chi^2 /{\it dof} = 0.37$, so a larger range of fit parameters
fall within the 68\% confidence limits in either case.

\INSERTFIG{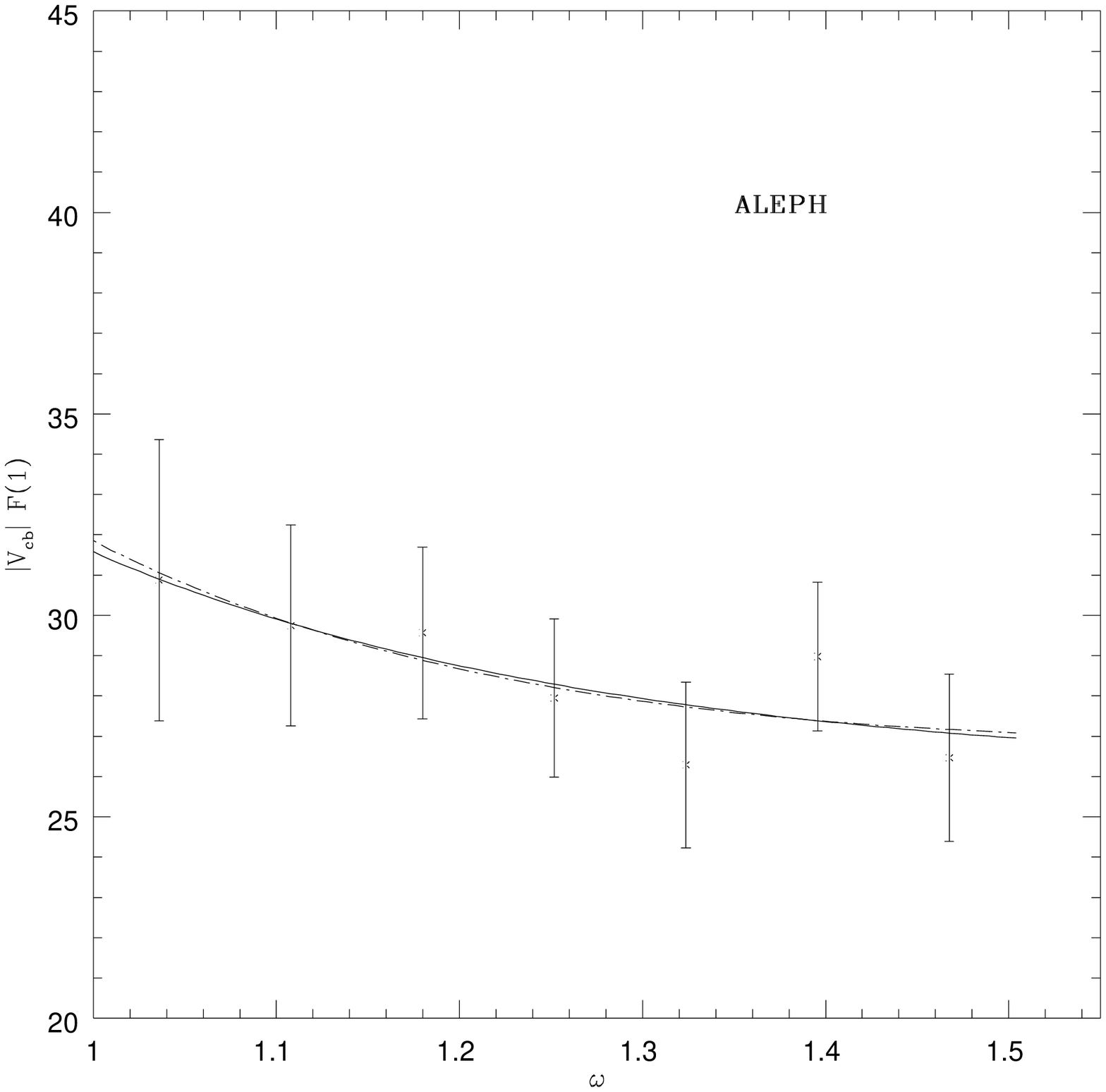}{2}{Fit of ALEPH data to our parametrization;
see Figure~1 caption for details.}

The constrained and unconstrained fits to ARGUS\argus\ data differ
mainly near zero recoil, with comparable values $\chi^2 /{\it dof} =
0.70$ and $0.67$, respectively (Fig.\ 3).  The ARGUS group used
several parametrizations, which yielded central values of $|V_{cb}|
\CF(1) \cdot 10^3$ from $39$ to $46$.
Their linear fit gave $|V_{cb}| \CF(1) \cdot 10^3 = 39 \pm 4$ and
$\CF'(1) = -1.17 \pm 0.11$, in some contrast to our central values.

\INSERTFIG{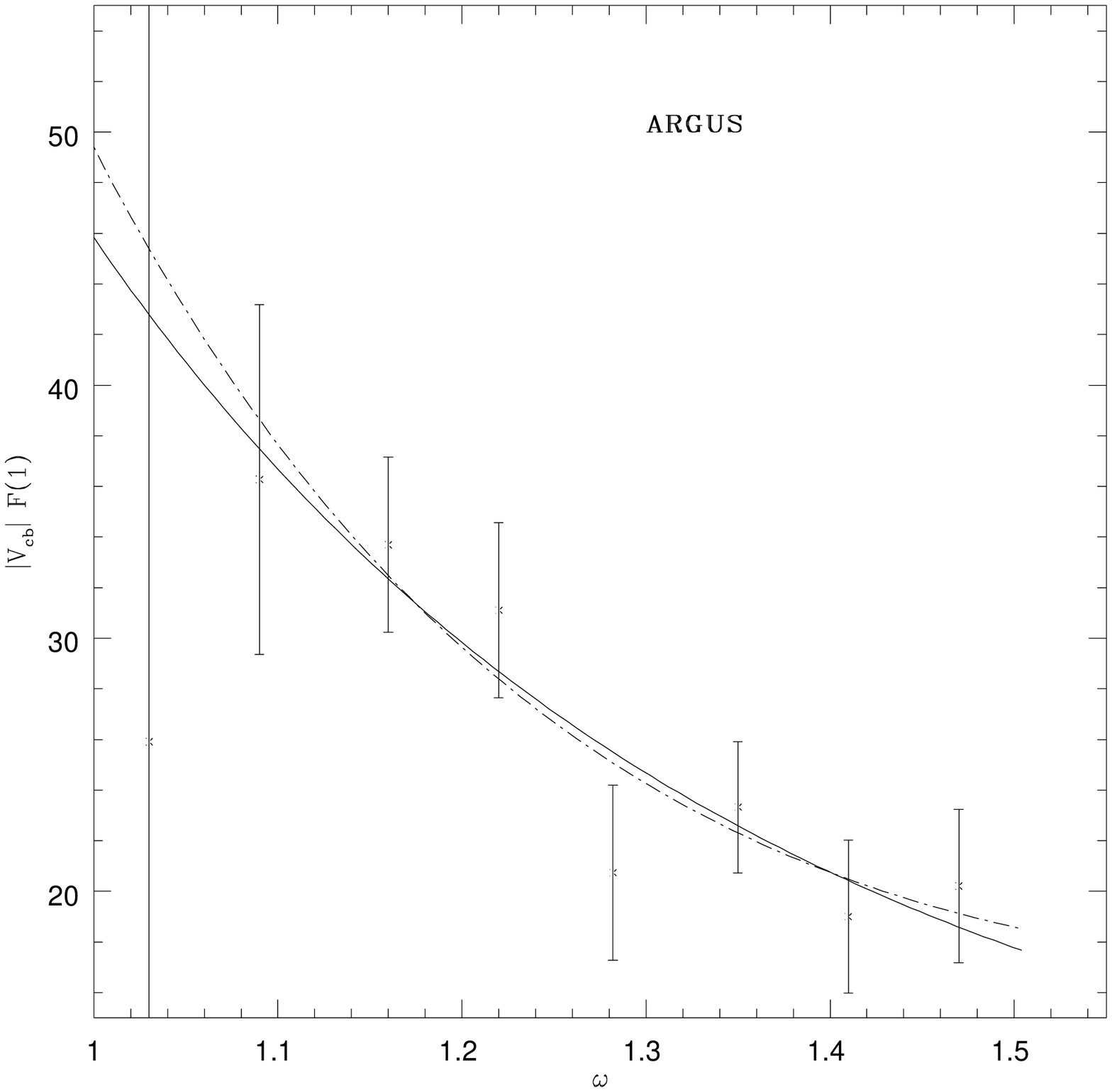}{3}{Fit of ARGUS data to our parametrization;
see Figure~1 caption for details.}

Although the large statistical uncertainty in $a_2/\CF(1)$ precludes
its determination at present, we can make a definite prediction for
the future: The central value of $a_2/\CF(1)$ must increase from
CLEO's present (unconstrained fit) number to fall inside our
bounds. Taking the theoretical estimates of Sec.\ 5 into account, we
predict $|a_2/\CF(1)| \leq 1.4 $.
 
\newsec{Conclusions}

Dispersion relation techniques and the use of analyticity properties
of hadronic form factors as functions of their kinematic variables
provide a valuable window into the realm of nonperturbative physics.
Using these methods, one can obtain useful bounds on quantities of
interest, in this case the form factors in the semileptonic decays
$\bar B \to D l \bar \nu$ and $\bar B \to D^* l \bar \nu$.

These bounds may be transformed into parametrizations of the four
experimentally accessible form factors relevant to $\bar B \to D l
\bar \nu$ and $\bar B \to D^* l \bar \nu$. Given the continuing
experimental scrutiny devoted to these decays, these form factors will
likely be measured in the foreseeable future.

Our derivation of these parametrizations relied on dispersion
relations, crossing symmetry, and a perturbative QCD calculation
performed at a scale $m_B + m_{D^*}$. The derivation improves on an
earlier work\bglapr\ in that no use of heavy quark symmetry was made.
The various uncertainties involved in the derivation, such as
perturbative corrections and uncertainties in quark masses, were
estimated, and shown to be unimportant. This includes effects from
branch cuts in the form factors due to non-resonant contributions.

The result is a three-parameter description of each of the form
factors $f_0$, $g$, and $f_+$ accurate over the entire physical
kinematic range to better than $2\%$. The value of one of the
parameters, the normalization at zero recoil $\CF(1)$, is predicted by
heavy quark symmetry.  The other two parameters $a_1, a_2$ are
constrained by $|a_1|^2 + |a_2|^2 \leq B^2$, with a leading-order
result $B=1$. A very conservative estimate of corrections to our
results leads us to conclude that to all orders, the bound obeys $B <
1.4$.  The $2\%$ or better accuracy of the three-parameter description
applies for any $B \leq 1.4 $. The three parameter fit to the form
factor $F_1$ is less accurate; for $B \leq 1.4 $ we find a bound of
$8\%$ on its relative error.

We emphasize that we have determined strict upper bounds on the
truncation errors. The truncation errors may be significantly
smaller. The strict inequality~\optical\ can be improved by including
the contributions of other intermediate states; our use of Blaschke
factors, Eq.~\blaschke, amounts to assuming the largest possible
uncertainty from the residues of poles in the form factors; and the
bounds on the parameters for each form factor, Eq.~\asum, are actually
correlated, as in Eqs.~\fg\ and~\ff.

As an application of our results, the individual form factors in $\bar
B \to D^* l \bar \nu$ were combined using heavy quark symmetries in
order to obtain a single parametrization of the differential
cross-section $d\Gamma/dq^2$, which was then fit to data.  This was
necessary because the best data currently available sums over $D^*$
mesons in all polarization states and thus involves more than one form
factor. However, to $\CO(1/m_c)$, our results depend only on charm
quark spin symmetry and the constant $\bar \Lambda = m_D - m_c$, and
are therefore expected to be more reliable than those using the full
flavor-spin heavy quark symmetry.  We obtain values for a
three-parameter ($|V_{cb}| \CF(1)$, $a_1/\CF(1)$, and $a_2/\CF(1)$)
fit to the single form factor $\CF(v \cdot v')$ describing $\bar B \to
D^* l \bar \nu$ that is free of the theoretical errors inherent in
choosing a parametrization for extrapolating the data to zero recoil.
We again emphasize that, although heavy quark spin symmetries were
used in obtaining values for $|V_{cb}| \CF(1)$, this is a limitation
imposed by the currently available data that will be lifted when
better measurements of the individual form factors become available.

The intensive experimental effort focused on semileptonic $\bar B \to
D l \bar \nu$ and $\bar B \to D^* l \bar \nu$ decays will result in
increasingly precise measurements of the rate and form factors. Our
descriptions of these form factors are remarkably insensitive to
theoretical uncertainties, and are accurate over the physical
kinematic range to better than $2\%$; as such, they should be useful
ingredients in studying the nonperturbative physics of semileptonic
$B$ decays.

\vskip1.2cm
{\it Acknowledgments}
\hfil\break
We would like to thank Hans Paar, Vivek Sharma, and Persis Drell for
useful discussions of the CLEO and ALEPH experiments.  The research of
one of us (B.G.)  is funded in part by the Alfred P. Sloan
Foundation. This work is supported in part by the Department of Energy
under contract DOE--FG03--90ER40546.
\vfill\eject
\listrefs
\bye